\documentclass[twocolumn,superscriptaddress,amsmath,amssymb]{revtex4}

\usepackage{bm}
\usepackage{amsmath}
\usepackage{amssymb} 
\usepackage{mathrsfs}
\usepackage{dcolumn}
\usepackage{indentfirst}
\usepackage{graphicx, makeidx}
\usepackage{epstopdf}
\usepackage{color} 

\begin{document}


\title{Aspiration of biological viscoelastic drops}        



\author {Karine Guevorkian}
\affiliation{
Unite Mixte de Recherche 168, Centre National de la Recherche Scientifique-Institut Curie, Paris 75248 cedex France
}

\author{Marie-Jos\'{e}e Colbert}
\affiliation{Department of Physics and Astronomy, McMaster University , Ontario L8S4L8, Canada}

\author{M\'{e}lanie Durth}
\affiliation{Ecole Polytechnique, Laboratoire d'Hydrodynamique, Palaiseau 91128 cedex, France}

\author{Sylvie Dufour}
\affiliation{Unite Mixte de Recherche 144, Centre National de la Recherche Scientifique-Institut Curie, Paris 75248 cedex, France}

\author{Fran\c{c}oise Brochard-Wyart}
\email{brochard@curie.fr}
\affiliation{
Unite Mixte de Recherche 168, Centre National de la Recherche Scientifique-Institut Curie, Paris 75248 cedex France
}



\begin{abstract}
Spherical cellular aggregates are in vitro systems to study the physical and biophysical properties of tissues. We present a novel approach to characterize  the mechanical properties of cellular aggregates  using micropipette aspiration technique. We observe an aspiration in two distinct regimes, a fast elastic deformation followed by a  viscous flow. We develop a model based on this viscoelastic behavior to deduce the surface tension, viscosity, and elastic modulus. A major result is the  increase of the surface tension with the applied force, interpreted as an effect of cellular mechanosensing. 

\end{abstract}

\pacs{87.18.R, 87.18.Ed, 87.18.Fx}
\keywords{cellular aggregates, biological tissue, micropipette aspiration, viscoelasticity}

\maketitle



Embryonic morphogenesis, wound healing, cancer growth and metastasis are a few examples where the physical laws play an important role along with genetic cues in the functioning of a tissue. An aggregate of living cells, used as a model tissue,  behaves like a viscoelastic liquid. Spreading and sorting are signatures of liquid-like behavior of embryonic tissues \cite{Gordon:72p43, Foty:1994p1001}. Moreover, cellular aggregates in solution round up to form ``spheroids'' in order to minimize their surface energy, similar to oil drops in water. This is a manifestation of surface tension, which has been related to intercellular adhesion energy \cite{Foty2005255}. In the past, the simple analogy between liquids and tissues has lead to valuable findings about the mechanics  of embryonic mutual envelopment \cite{Foty:1996p967}, tissue spreading \cite{Ryan:2001p969}, and cancer propagation \cite{Foty:cancer}. A knowledge of the surface tension of tissues has also been essential for organ printing in tissue engineering \cite{Jakab:2008p413}.

To measure the surface tension of cellular aggregates and investigate the role of surface tension in cell sorting, Steinberg and coworkers \cite{Foty:1994p1001} introduced the parallel plate compression apparatus, which has since been used by other groups \cite{Norotte:2008p1000, Mgharbel:2009p952}. In this method, an aggregate is subjected to an imposed deformation and the surface tension is inferred from the relaxation force, while the viscosity of the tissue is obtained from the shape relaxation  \cite{Mombach:2005p1003}. Difficulties in the evaluation of the principal radii of a compressed aggregate and the contact angle between the aggregate and the plate make this technique rather delicate. Deformation of aggregates under centrifugal forces is an alternative way that has been used to classify aggregates of various cell types \cite{phil69}. Recently this technique has been combined with Axisymmetric drop shape analysis (ASDA) for measuring the surface tension of embryonic tissue \cite{Kalantarian:2009p1161}.

In this letter, we propose the use of micropipette aspiration technique to study the surface tension and the mechanical properties of cellular aggregates. This technique has previously been used to evaluate the viscoelastic properties of single cells \cite{Sato:1990p1004, Evans:1989p200} and the stiffness of  tissues \cite{Aoki:1997p985, Butcher:2007p1002, ToshiroOhashi:2005p1160} at small deformations. For a Newtonian fluid, the aspiration dynamics is governed by the Washburn law, $L(t)\sim t^{1/2}$, where $L(t)$ is the advancement of the liquid inside the pipette \cite{wash}. For a tissue, a completely different behavior is observed due to its viscoelastic properties. Under applied stress $\sigma$, a tissue responds like an elastic solid at times shorter than a characteristic time $\tau$ \cite{Chu:1975p984}, and like a fluid for $t>\tau$. This behavior can be described by $d\sigma/dt+\sigma/\tau=E d\epsilon/dt$, where $\epsilon$ is the strain; the viscosity $\eta$, of the material is related to its elastic modulus $E$, through $\eta\approx E\tau$ \cite{landau}. In the case of parallel plate compression, $\epsilon$ is constant and the stress relaxes to equilibrium, whereas for the case of aspiration, $\sigma$ stays constant and the tissue flows. 

Spherical cellular aggregates are useful systems to study the mechanical properties of tissues since the adhesion energy between the subunits (cells) can be controlled. We have used murin sarcoma (S180) cell lines transfected to express various levels of E-cadherin molecules at the surface of the cells \cite{Chu:2004p968}, thereby controlling the intercellular adhesion energy. Here, we focus on the most adhesive cell lines. Cells were cultured under 5$\%$ air/ 5$\%$ CO$_2$ atmosphere in DMEM enriched with 10$\%$ calf serum (culture 
medium) and prepared for aggregation following a procedure similar to Ryan et al.'s  \cite{Ryan:2001p969}. Aggregates ranging from 250 $\mu$m to 400 $\mu$m in diameter were obtained from 5 ml of cell suspension in CO$_2$-equilibrated culture medium at a concentration of $4\times 10^5$ cells per ml in 25 ml erlenmeyer flasks, and placed in a gyratory shaker at 75 rpm at 37$^\circ$C for 24 hours. The flasks were pretreated with 2$\%$ dimethylchlorosilane in chloroform and coated with silicon to prevent adhesion of cells to the glass surface. We performed the aspiration of the aggregates using pipettes with diameters 3-5 times that of a single cell (40-70 $\mu$m). The pipettes were fabricated by pulling borosilicate capillaries ($1$ mm/$0.5$ mm O/I diameter) with a laser-based puller (P-2000, Sutter Inst. Co, Novato, CA ), and sized to the desired diameter by using a quartz tile. To prevent adhesion of the cells to the micropipette walls, the pipettes were incubated in 0.1 mg/ml PolyEthyleneGlycol-PolyLysin (PLL(20)-g[3.5]-PEG(2), Surface Solution, Dubendorf Switzerland) in HEPES solution (pH 7.3) for one hour. The observation chamber consisted of a thick U-shaped Parafilm spacer (2 cm$\times$2 cm$\times$5 mm), sandwiched in between two microscope slides by gentle heating. Aggregates were then suspended in CO$_2$ equilibrated culture medium and the pipette was introduced into the chamber. To prevent evaporation, the open end was sealed with mineral oil. A large range of pressures ($\Delta P=0.1-5$ kPa) was attained by vertically displacing a water reservoir, connected to the pipette, with respect to the observation chamber. Aspirated aggregates were visualized on an inverted microscope (Zeiss Axiovert 100) equipped with a $\times$20 air objective (NA 0.45). Movies of the advancement of the aggregates inside the pipette were recorded with a CCD camera (Luca-R, Andor, Belfast UK) with a 5-30 second interval. Cell viability in aspirated aggregates was checked using trypan blue exclusion test. After 3 hours of aspiration, trypan blue was added to the experimental chamber to a final concentartion of $25\%$. A small number of dead cell were present at the core of the aggregate, comparable to the aggregates at rest, but no significant cell death was seen in the aspirated tongue.

\begin{figure}
 \centering
 \includegraphics[width=3in]{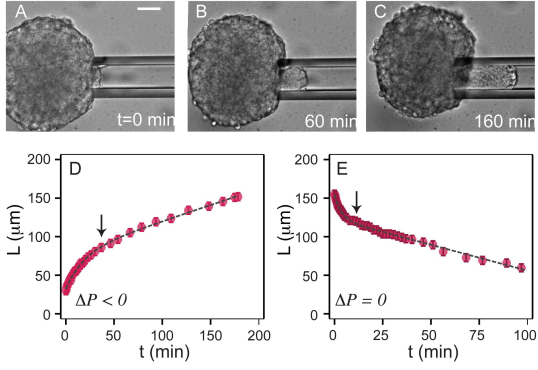} 
   \caption{Micropipette aspiration of spherical cellular aggregates. (A)-(C) Aspiration of an aggregate with $\Delta P=1370$ (14 cmH$_2$O), $R_0=150$ $\mu$m, $R_p=30$ $\mu$m, scale bar is 50 $\mu$m. (D) Aspiration and (E) retraction cycles for an aggregate at  $\Delta P=1180$ Pa, with $R_0=175$ $\mu$m, and $R_p=35$ $\mu$m. Arrows indicate the transitions from elastic to viscous regimes. Dotted lines are fits to the experimental curves using the viscoelastic model (see text for details). }
 \label{fig_1}
\end{figure}
\begin{figure}
 \centering
 \includegraphics[width=3.25in]{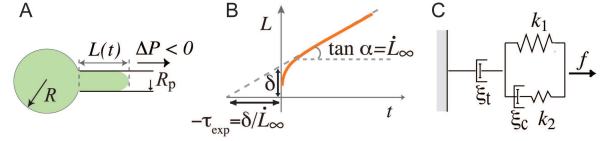} 
   \caption{Aspiration of a viscoelastic drop. (A) Schematic presentation of an aspirated drop. (B) Creep curve showing a fast elastic deformation, $\delta$, followed by a viscous flow, $\dot{L}_\infty$. (C) Modified Maxwell model. The Kelvin body accounts for the initial elastic deformation, where $k_1$ is the spring constant related to the elasticity of the aggregate, $k_2$ accounts for the initial jump in $L(t)$, and $\xi_c$ is a local friction coefficient, related to the raising time of the elastic deformation. The dashpot represents the viscous dissipation of the flowing tissue.}
\label{fig_2}
\end{figure}

Fig.\ref{fig_1}(A)-(C) shows snapshots of the aspiration of an aggregate inside a pipette at a constant pressure. The advancement of the aggregate inside the pipette is characterized by tracking the displacement of the front of the tongue with respect to the pipette tip, represented by $L(t)$ in Fig. \ref{fig_2}(A). As a first approach, steps of $\Delta P$ were applied at a time interval of 2-3 hours, in order to determine the dynamics of aspiration as a function of $\Delta P$. However, we observed a degradation of the cells when aggregates stayed under aspiration for over 6 hours, limiting the number of steps. Consequently, we modified the procedure and applied cycles of pressure as shown in Fig. \ref{fig_1}(D)-(E). After each aspiration at constant pressure, the pressure was set to zero and the retraction of the tongue was monitored. In general, we performed one aspiration-retraction cycle on each aggregate to maintain the same initial conditions. Both aspiration and retraction curves show a fast initial deformation, followed by a slow flow with constant velocity $\dot{L}_{\infty}$; the transition between the two regimes is marked by an arrow. This creep behavior is a signature of viscoelastic materials.  We proceeded by considering these cell aggregates as viscoelastic liquid drops with a surface tension $\gamma$.

The total energy of a drop aspirated inside  a non-adhesive pipette, ``zero'' wetting, is given by $\mathcal{F} = (4\pi R^2 + 2\pi R_p L)\gamma-\pi R_p^2L\Delta P$ \cite{degennes:2004}, where $R$ and $R_p$ are the radii of the drop and the pipette respectively, and  $\Delta P$ is the applied pressure, as shown schematically in Fig. \ref{fig_2}(A). Considering volume conservation,  the aspiration force is $ f =\pi R_p^2 ( \Delta P-\Delta P_c)$, where the critical pressure to aspirate, $\Delta P_c$, relates to the surface tension through the Laplace law: $\Delta P_c = 2\gamma(\frac{1}{R_p}-\frac{1}{R})$. Note that $R$ is not constant, but can be approximated by $R\approx R_0$ for $R_p\ll R_0$.
From scaling laws the aspiration force and the elastic deformation at short time, $\delta$, are related by $\frac{f}{A_0}=C\times E\frac{\delta}{R_p}$, where $A_0=\pi R_p^2$, $E$ is the elastic modulus, and $C$ is a geometrical factor  $C\approx1$ for our experimental conditions \cite{Aoki:1997p985},  leading to $f\approx\pi R_p E \delta$. At long times, $f$ is balanced by the friction force due to the viscous flow into the orifice \cite{Dagan:1982p1007} and the slippage of the advancing tongue on the wall as: $f=3\pi^2\eta R_p\dot{L}+2\pi kR_pL\dot{L}$, where $\eta$ is the viscosity of the tissue, and $k$ is the wall-tissue friction coefficient. We define $L_c=3\pi\eta/2k$ as a characteristic length associated to the wall friction. In the limit of $L>L_c$ we have $L\sim t^{1/2}$, whereas for $L<L_c$ we find $L=\dot{L}_\infty t$, where $\dot{L}_\infty=f/3\pi^2\eta R_p$. We have estimated $k\approx 10^8$ N.s/m$^2$ from the advancement velocity of a completely aspirated aggregate, leading to $L_c\approx \eta/k \approx 2$ mm (see below for $\eta$). Therefore we can ignore the wall friction. To combine the elastic and viscous regimes, we use the modified Maxwell model depicted in Fig. \ref{fig_2}(C). The total displacement $L(t)$ is given by:
\begin{equation}\label{eq:L}
L(t)=\frac{f}{k_1}\left(1-\frac{k_2}{k_1+k_2}e^{-t/\tau_c}\right)+\frac{f}{\xi_t}t,
\end{equation}
where $k_1=\pi R_p E$, and $\xi_t=3\pi^2\eta R_p$. The first term characterizes  the elastic regime with $\tau_c=\frac{\xi_c(k_1+k_2)}{k_1k_2}$ being the raising time of the elastic deformation $\delta$, and the second term characterizes the flow at constant velocity $\dot{L}_\infty$. The tissue relaxation time separating the elastic and viscous regimes is given by $\tau=\xi_t/k_1=3\pi\eta/E$.

\begin{figure}
 \centering
 \includegraphics[width=3.3in]{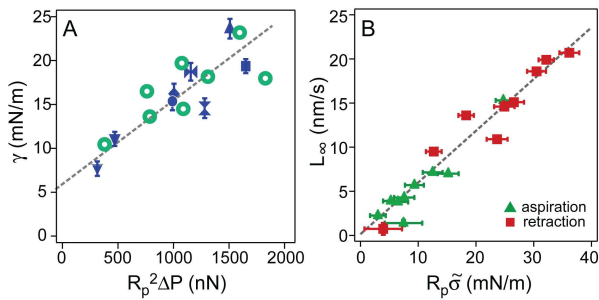} 
   \caption{Viscosity and surface tension of aspirated aggregates. (A) $\gamma$ as a function of applied force $R_p^2\Delta P$. Filled symbols are obtained from the relationship between $\dot{L}^a_\infty$ and $\dot{L}^r_\infty$ and open symbols are obtained from $\dot{L}^a_\infty$ and using the measured value for $\eta$. The curve is presented to guide the eye.  (B) Flow velocity $\dot{L}_\infty$ as a function of $R_p\tilde{\sigma}$, (slope equals to $1/3\pi\eta$).}
 \label{fig_3}
\end{figure}

The dashed lines in Fig. \ref{fig_1}(D)-(E) are the adjustment of Eq. \ref{eq:L} to the data with four fitting parameters: $\delta=f/k_1$, $\dot{L}_\infty=f/\xi_t$, $\beta=k_2/(k_1+k_2)$, and $ \tau_c$. The critical pressure is deduced from $\Delta P_c=\Delta P\dot{L}^r_\infty /(\dot{L}^r_\infty+\dot{L}^a_\infty)$, where $\dot{L}^a_\infty$ and $\dot{L}^r_\infty$ are the aspiration and the retraction flow rates, respectively. Using the values for $\Delta P_c$, the surface tension, $\gamma$, is derived from the Laplace law. Fig. \ref{fig_3}(A) shows an increase in $\gamma$ as the applied force is increased. By extrapolation, we obtain the surface tension of the aggregate at rest, $\gamma_0\approx 6$ mN/m, comparable to previously obtained values for similar tissue types \cite{Foty:1994p1001, Forgacs:1998p958, Mgharbel:2009p952}. We also measured directly a lower bound for $\Delta P_c (\gamma_0)$ by finding the maximum pressure ($\approx 300-400$ Pa for $R_p=35$ $\mu$m) at which the aggregate does not penetrate into the pipette, leading to  $\gamma_0 \sim 5-7$ mN/m.
The flow velocities of aggregates during aspiration and retraction are shown in Fig. \ref{fig_3}(B) as a function of the applied stress, $\tilde{\sigma}$, where  $\tilde{\sigma}=\Delta P -\Delta P_c$ for aspiration, and $\tilde{\sigma}=\Delta P_c$ for retraction. The observed linear relationship between $\dot{L}_\infty$ and $R_p\tilde{\sigma}$ shows that $\eta$ stays constant and no shear thinning effect is observed in the range of pressures (1-3 kPa) applied in our experiments. The slope of the  fitted line gives $\eta=1.9\pm0.3 \times 10^5$ Pa.s,  comparable to the values previously reported for aggregates of mouse embryonal carcinoma F9 cell lines ($\eta\approx 2\times 10^5$ Pa.s) \cite{Mgharbel:2009p952, Marmottant:2009p1005}  and various chicken embryonic tissues ($\eta\approx \times 10^5$ Pa.s) \cite{Forgacs:1998p958, Jakab:2008p953}. Preliminary results on aggregates of the same cell lines with less intercellular cohesion have shown a similar but much faster aspiration dynamics, indicating a smaller viscosity for these aggregates (data not shown). In our analysis we have assumed that $\Delta P_c$ does not relax in the time scale of our experiment when $\Delta P=0$. This assumption is justified, since the slopes of the fast retraction curves stay constant as seen in Fig. \ref{fig_1}(E).

As mentioned above, the relaxation time for a viscoelastic material to flow is $\tau\approx\eta/E$. This characteristic time can experimentally be evaluated from the creep curve as $\tau_{exp}=\delta/\dot{L}_\infty$ (Fig. \ref{fig_2}(B)). However, as can be seen from the curves on Fig. \ref{fig_1}(D)-(E), the retraction of the tongue has a much faster dynamics, resulting in $\tau_{exp}^a\gg \tau_{exp}^r$. This is due to  $\gamma$ increasing from $\gamma_0$ (elastic regime) to $\gamma$ (viscous regime) during the slow aspiration and not relaxing during the fast retraction. Taking these corrections into account, $\tau=\tau_{exp} \times f_{visc}/f_{elastic}$, leading to $\tau^a=\tau_{exp}^a\frac{\Delta P-\Delta P_c(\gamma)}{\Delta P-\Delta P_c(\gamma_0)}$, and $\tau^r=\tau_{exp}^r\frac{\Delta P_c(\gamma)}{\Delta P}$. Taking $\gamma_0=6$ mN/m we obtain $\tau^a=47\pm10$ min., and $\tau^r=40\pm7$ min., resulting in an average value of $\bar\tau=44\pm7$ min. We estimate an elastic modulus of $E= 3\pi\eta/\bar\tau\approx700\pm100$ Pa for these aggregates, comparable to values reported for embryonic liver tissue \cite{Forgacs:1998p958}. The elastic local cell's relaxation time, $\tau_c$, is one order of magnitude smaller than the tissue relaxation times. We systematically find $\tau_c^a\gg\tau_c^r$, showing that pre-stressed tissue has a faster elastic response.
\begin{figure}
 \centering
 \includegraphics[width=3in]{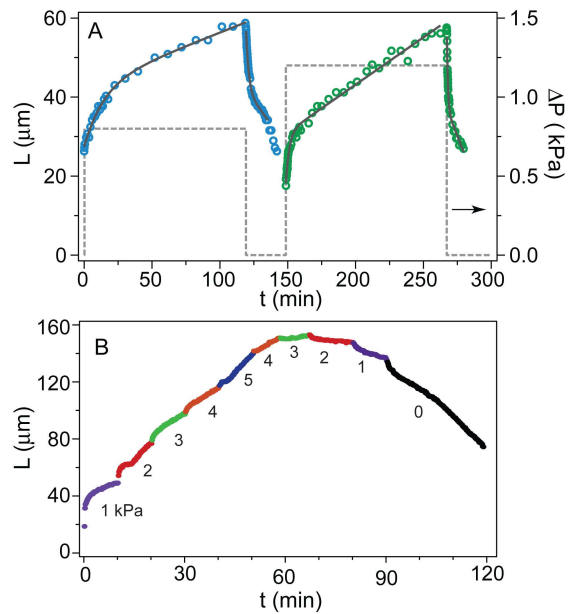} 
   \caption{Manifestation of surface tension augmentation. (A) Successive aspiration on the same aggregate. Dashed lines represent the pressure profile. The second aspiration shows a smaller elastic deformation in spite of the applied $\Delta P$ being larger . (B) Increasing and decreasing $\Delta P$ in steps. When $\Delta P$ is decreased to 3 kPa, the aggregate stops flowing, indicating that the surface tension has increased from its steady state value.}
 \label{fig_4}
\end{figure}
We have characterized mechanical properties of tissue such as their surface tension, viscosity and elasticity using micropipette aspiration technique. We have found that the surface tension of the aggregate is stress-dependent, suggesting that upon the application of a permanent external force, tissue cohesion is reinforced. Successive aspiration on the same aggregate validates our finding. As shown in Fig.\ref{fig_4}(A), the elastic deformation of the second aspiration in smaller, indicating a larger initial $\gamma_0$. Another direct manifestation of the reinforcement of $\gamma$ is shown in Fig.\ref{fig_4}(B), when $\Delta P$ is decreased  to a few times $\Delta P_c(\gamma_0)$, the aggregate relaxes instead of flowing.

The reinforcement of $\gamma$ is a signature of an active response of the cells to mechanical forces \cite{Cai200947, Janmey2004364} leading to cytoskeletal remodeling \cite{nborghi2009}, which may involve stretch-activated membrane channels \cite{Sbrana:2008p1199}, stress fiber polymerization and  tensening by Myosin II motors \cite{Chaudhuri:2009p1201, Desprat:2005p1203}, and clustering of cadherins \cite{DelanoeAyari:2004p1200}. At the tissue level, it has also been shown that application of an external force to the tissue using a 20 $\mu$m micro-needle increases the tissue tension, leading to morphogenetic movements \cite{Pouille04142009}. Protein labeling and cytoskeleton modifying drugs have to be used to better understand the reinforcement mechanism at a the cellular level. This novel method brings complementary features to the classical parallel plate compression technique, since instead of relaxing to equilibrium, the cells flowing into the pipette are continuously stretched. Moreover, this technique allows us to reach much higher stresses, up to hundred times the aggregate's Laplace pressure.

How the surface tension and the viscoelastic properties of an aggregate depend on the properties of the subunits and on their interconnection remains an open question. Previous studies have measured the surface tension of aggregates as a function of the level of expression of intercellular binders (cadherin molecules) \cite{Foty2005255}. However, the relationship between the adhesion energy and the surface tension is still debated. We anticipate using the micropipette aspiration technique to relate the surface tension of aggregates to the cell-cell adhesion energy, which has been previously measured  by one of us \cite{Chu:2004p968}. 

Complete aspiration of aggregates inside a pipette can also be used to apply high pressures ($\sim\gamma/R_p$) to cancerous tissue and thus investigate the validity of the homeostatic pressure model, which predicts that metastatic cells can only grow if the internal pressure of the aggregate is below a critical ``homeostatic pressure''  \cite {Basan:2009p1006}. Combined with confocal microscopy, tissue relaxation under stress can be studied at microscopic level by probing the cellular rearrangements inside an aspirated aggregate. Compared to more conventional methods, the micropipette aspiration technique is easy to set up and can be applied to in-vivo examination of biological systems, such as living tissue or drug treated tumors, and to other complex fluids, such as viscous pastes and foams.

We would like to thank D. Cuvelier for his help with the experimental setup, C. Clanet for useful discussions, and J. Elgeti and D. Gonzalez-Rodriguez for their critical reading of the manuscript. F. B. W. and S. D. would like to thank Curie PIC program for funding. The group belongs to the CNRS consortium CellTiss.  

%

\end{document}